# AI-powered Augmented Reality as a Threat Vector for Human Manipulation

**Louis Rosenberg**, PhD
**Unanimous AI**
**Arlington, Virginia, USA**

## Abstract

Augmented Reality (AR) is a powerful perceptual technology that can alter what users see, hear, feel, and experience throughout their daily lives. When combined with the speed and flexibility of context-aware generative AI, the power is greatly expanded, allowing individual users to be targeted with custom-generated AR experiences that are instantly tailored to who they are, where they are, and what they are doing. This can transform the physical world into a magical place, but only if the augmentation of a user's environment is enacted for their personal benefit and best interests. Instead, if AI-powered AR systems are controlled by unregulated third parties, such as large corporations or state actors, individually adaptive AR experiences could be deployed as a dangerous form of targeted influence. In fact, if the industry adopts an advertising business model for AI-powered AR devices, context-aware *generative influence* could become a widely used manipulative path for promotion of products and services in the physical world. Worse, similar techniques could be used for political influence, propaganda, and disinformation. This chapter reviews the power and flexibility of AI-generated augmented reality, explores the risks that emerge when used for persuasion, manipulation, or influence, and proposes policy directions to mitigate these risks.

## 1. Introduction

Augmented Reality (AR) is a powerful technology that merges realistic virtual elements into a user's experience of the real physical world [1,2]. It is often confused with simpler technologies such as smart glasses and head-up displays that merely annotate or embellish a user's field of view, but do not actually alter the user's mental model of reality. True AR (also often referred to as mixed reality) is a transformative cognitive experience in which the virtual content is so naturally and seamlessly integrated into a user's perception of a real physical space that their brain builds *a single mental model* of their ambient environment that incorporates both real and virtual elements perceived as a single unified reality [3-6].

The distinction between annotating a user's view and augmenting their reality is important because true AR is a highly impactful experience in which the user can easily confuse what is real and what is computer-generated. This has the potential to unleash amazing applications across a wide range of fields from entertainment and education to medicine, commerce, science and engineering. It also has the potential to be an extremely dangerous vector for targeted influence, especially when third parties such as corporations or state actors can selectively alter a user's perception of their physical surroundings as they navigate their daily lives [7].

Because the difference between smart glasses and true augmented reality has a profound impact on the benefits and risks of the experiences delivered, it is helpful to identify the cognitive process that governs whether an interface augments a user's reality (i.e., the virtual content is authentically integrated into their mental model of reality) or simply annotates their field of view. To make this rigorous, we must introduce the psychophysical concept of **Distal Attribution** [8]. It refers to our brain's ability to receive a piece of perceptual content (e.g., sight, sound, touch, smell, heat, etc.) and "externalize it" such that we experience the perceptual content as an authentic aspect or element of the real world around us [5,10].

This process is so fundamental to our sense of reality that we rarely think about its unique importance. To explore, consider this simple thought experiment – you are holding a walking stick as you climb a trail in a woody forest. You place the stick down and it hits a soft patch of mud just one step ahead of you. Your brain quickly updates its mental model of your surroundings and adjusts your gait for the soft ground your foot is about to land on. You proceed smoothly without a stutter in your step, all thanks to your rapidly updated model of reality.

This all seems obvious until you realize that the sensation your brain received was not imparted by the soft, muddy ground – it was imparted by the walking stick in your hand. From your brain's perspective, nerves fired in your muscles as a result of the walking stick impacting your body through your hand. Your brain could have interpreted these signals as the walking stick suddenly going soft, suggesting it can't support your weight anymore. But that's not what happened – your brain *distally attributed* the perceptual information to the ground on the other end of the stick. In other words, your brain correctly projected the sensory information into an external spatial model of the world. That model is what you perceive as reality.

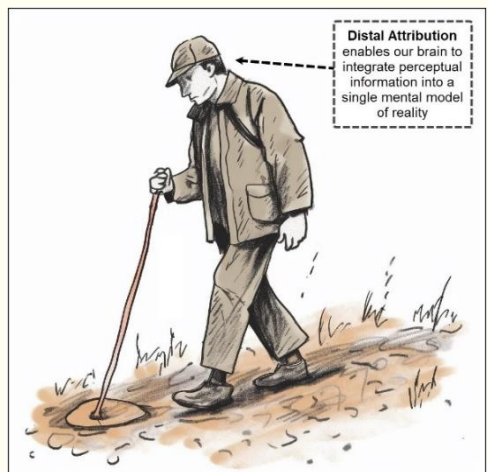

Fig 1. Distal Attribution turns perceptions into our mental model of reality.

This example highlights the fact that the reality we experience is not a direct representation of the sensory information reaching our brains but the result of a complex process in which sensory content is incorporated into an internal mental model we experience as reality. It is not deterministic, meaning perceptual content can either (a) be correctly added to our mental model of reality, (b) be incorrectly added to our mental model of reality, or (c) simply fail to integrate with our mental model at all. If you project a standard movie on a wall, it could depict people in an adjacent room, but you're unlikely to believe the scene extends into the wall. That's because it does not get integrated into your mental model of reality. It's perceived as an annotation on top of reality and is analogous to the experiences provided by early head-up displays [9] and today's smart glasses and weak AR devices.

Distal Attribution rarely fails in daily life because what we see, hear, feel, and experience are usually highly consistent, both spatially and temporally [11-13]. In fact, the only breakdown of distal attribution that I regularly experience in the real world is when I get my car washed. Despite knowing that my car is parked and that large furry rollers are moving outside my windows, my brain often builds a mental model in which I actually feel my car moving. This error is caused by a mismatch between my vestibular and visual senses. My brain is forced to choose between a model of reality where either my car is moving forward or the scene outside my windows is moving backwards. This choice is called a *Perceptual Hypothesis* and is influenced by my senses in combination with my memories of prior experiences driving in cars. My memories win out and my brain makes the wrong hypothesis.

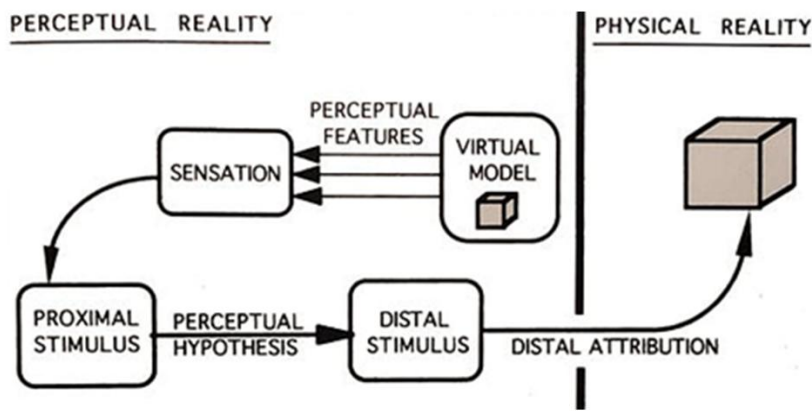

Fig 2. Diagram of Distal Attribution in Augmented Reality (Reprint, Rosenberg 1994)

Because Distal Attribution is the process by which our brain integrates content into our mental model of reality, achieving it is central to enabling true augmented reality experiences [5, 10-13]. The first AR system to foster distal attribution and thereby enable users to experience a single *unified reality* of real and virtual objects was the Virtual Fixtures Platform developed at the Air Force Research Laboratory (1991–1994) [3, 5, 14]. That early research uncovered guidelines for interactive virtual objects to be perceived as believable additions to a user's "ambient reality." These requirements for generating an authentic mixed reality experience include: (a) real and virtual objects must be spatially registered in 3D within the limits of human perception, (b) the user must be able to interact naturally with both real and virtual objects, (c) real and virtual objects must interact with each other in authentic ways, and (d) the sensory modes of sight, sound, touch, and proprioception must be synchronized in time within human perceptual limits. For example, it was found during the Virtual Fixture studies that a relative time delay as low as 70 ms among sight, sound, and touch could hinder distal attribution and make interactive virtual objects not be accepted as authentic elements of a real environment [16].

In addition, this research found that "a natural and predictable relation" must exist between *efference* (i.e., a user's active engagement with an environment) and *afference* (i.e., that user's changing perception of the environment) [5]. This might sound obvious, but it drives one of the most common failure modes that hinders distal attribution in AR and breaks suspension of disbelief.  Consider a user placing a virtual book on a real desk.  If the virtual book sinks into the real desk, afference and efference do not match, and the book may not be integrated into the user's mental model of reality. If the solution is to restrict penetration of the virtual book into the real desk as a visual illusion, the book might appear on the desk, but the user's hand will likely have penetrated below the desk (as perceived proprioceptively), which also creates a perceptual mismatch that hinders AR by breaking distal attribution.

To address such a mismatch,  a variety of techniques have been developed. For example, haptic "impulse" sensations that mimic the instant of contact between the virtual book and the physical desk have been found to mask proprioceptive conflict and sustain distal attribution [17].  Other techniques that help ensure virtual objects are perceived as authentic elements in real environments include the casting of virtual shadows on real surfaces and the simulation of real-world lighting conditions on virtual objects [18, 40].  Such cues have been found to increase realism, foster distal attribution, and enhance depth perception [19].  Also, "Sound Field Matching" in the audio domain can ensure that virtual sounds match the acoustics of real environments [41].  In addition, auditory cues that emulate interactions between virtual and real objects can significantly enhance realism and performance [20, 49]. And finally, pinning virtual objects to physical artifacts for tangible manipulation can support realism by aligning visual, haptic, and proprioceptive senses [21].

What all of these examples have in common is that enabling users to form a unified mental model of reality that integrates both real and virtual content requires achieving sufficient levels of *perceptual consistency* (i.e. spatial registration, temporal synchronization, and predictable relations between efference and afference among real and virtual objects). What is *not* required is perfect visual, audio, or haptic fidelity. That's because our brains require consistent percepts, not high-fidelity percepts, in order to accept elements into our reality. For example, if you walk into a dark room where you can barely see the furniture,  that room is just as real to you as it is under bright lights. However, if you bump into a table when crossing that dark room and your leg passes through it, realism is instantly lost.

### 1.1 True Augmented Reality is Finally Here

As described above, it is challenging to create perceptual experiences in which the real and virtual are genuinely perceived as authentic elements of a single unified reality. This can be referred to as *true augmented reality* or Mixed Reality [15, 37, 38, 45, 47]. For decades, this has only been possible in research labs and highly controlled product demos, but recently two high profile devices, the Apple Vision Pro and Meta Quest 3 were released that finally cross the "uncanny valley" of augmented reality at consumer scale. The Apple Vision Pro achieves this by accurately registering virtual content to the physical environment using LiDAR, depth sensors, and high-resolution cameras that work together to map the user's surroundings and identify real surfaces as spatial anchors.  It then uses a combination of precision head-tracking, pass-through video, and AI processing to maintain the registration and stability of virtual objects in all six degrees of freedom. Powered by Apple's custom M2 and R1 chips, the system achieves AR with very low latency, ensuring that sights, sounds, and motions maintain spatial congruence and temporal synchrony. The result is a transformative experience for many users [22].

Similarly, the Meta Quest 3 also employs sophisticated depth sensors and high-resolution cameras to map and recognize real-world surfaces, objects, and spatial layouts in real-time. Inside-out tracking enables alignment and stability of virtual elements as the user moves, while hand-tracking enables natural interaction. Powered by the Snapdragon XR2 Gen 2 chip, the system renders content with very low latency, maintaining coherence and fostering distal attribution. In addition, the Meta Quest 3 provides improved haptic feedback compared to their prior systems, enabling more subtle sensations as real and virtual elements interact.

While the Vision Pro and Meta Quest 3 represent major strides in the production of high-quality headsets, even bigger advancements were made on the software side in 2024 that may finally make AR devices an essential aspect of our digital lives, potentially replacing the mobile phone as our primary interface within the next five years. The core advancement was the widespread deployment of next-generation AI systems called Multimodal Large Language Models (MLLMs) which enable foundational AI systems to receive as input, not just text, but real-time streams of video and audio. This means that AI assistants and other onboard AI systems have "eyes and ears" that allow them to react to the context of real-world environments.

When MLLMs are integrated into augmented reality eyewear that contain onboard cameras and microphones, AI assistants and other generative models powering the experience can now see and hear everything the user sees and hears in real-time. The addition of context-aware AI agents to AR systems means that every user will be able to navigate their world with an AI assistant *hovering by their shoulder*, scanning their surroundings in real-time, and alerting them to things of interest, informing them of factual information, reminding them of nearby tasks, and providing them with custom-generated AR content in any place at any time. This, more than any other advancement in the evolution of AR, has the potential to mass-produce AR experiences for consumers that feel like superpowers [23, 25].

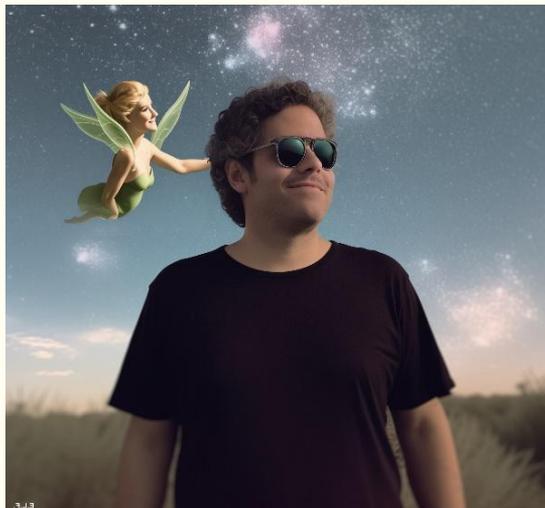

Fig 3. Our augmented future may include "ELFs" whispering in our ears.

These context-aware AI agents will provide real-time information via eyewear through text and imagery, but even more importantly they will assist users through natural dialog delivered through earbuds. This means that in the near future, we may all find ourselves with an AI whispering in our ears as we walk down the street, browse store shelves, hike mountain paths, attend classes, or perform jobs. This immersive "whisperverse" will feel like a mental superpower – a voice in our heads that continually informs us, reminds us, guides us, and even entertains us [24].

In some embodiments, these AI assistants will be deployed as *"Electronic Life Facilitators"* (or *ELFs*) that appear visually to us in augmented worlds as embodied characters. As shown in Figure 3, I believe these AI-powered entities will likely be rendered in AR as cute creatures that earn our trust, hovering around us as they give guidance or flying ahead to point out elements in our world [25, 26]. Whether deployed as disembodied voices or personified characters, the addition of context-aware AI assistants to our augmented future will be so impactful to our daily lives, I refer to this capability as *Augmented Mentality* and predict it will drive mass adoption [27]. After all, anyone not wearing AI-powered AR eyewear will be at a cognitive disadvantage.

The possibility of mass-produced AI powered AR glasses may have seemed out of reach only a few years ago, but over the last six months Meta and Google have both unveiled high-profile projects to bring these capabilities to consumers. Meta has already deployed Multimodal Large Language Models for use in Meta Ray-Ban smart glasses and Meta Quest 3 headsets. In December 2024, Google formally announced their long-rumored Project Moohan aimed at deploying AI-powered glasses and headsets in partnership Samsung, driven by a spatial operating system now called Android XR. In other words, Meta and Google are now racing to commercialize at scale, the superpowers that AI-enabled AR glasses will provide.

In addition to adding powerful assistive features, context-aware AI has the potential to make AR experiences significantly more realistic than ever before. As described above, the cognitive process of *distal attribution* occurs when AR content is so naturally integrated into our perceived surroundings, it gets incorporated into our mental model of reality. Context-aware AI will enable AR content to instantly adapt to changing conditions of our surroundings. This will reduce perceptual inconsistencies between the real and virtual, fostering distal attribution. In addition, when AI assistants are added to AR worlds that can see and hear everything that we are experiencing in real-time, these simulated entities will be significantly more likely to be accepted as authentic participants in our reality. Even if these agents are not truly intelligent, they are more likely to be perceived as "conscious beings" if they are responsive to our shared environment. This cognitive process is often called "Anthropomorphic Attribution" [28] and is likely to make context-aware AI assistants in AR seem far more real to us than those on flat screens [29].

## 3. AI-powered Augmented Reality will be Magical but Dangerous

With large corporations such as Meta, Apple, Google, and Samsung racing to deploy AI-powered eyewear that can provide convincing mixed reality experiences, our daily lives may soon be filled with immersive digital content that we engage all around us. This will enable magical applications for consumers and professionals. It will also unleash new forms of targeted influence ranging from immersive ads to new tactics for fraud, manipulation, and deception [14, 29, 30, 48]. Specifically, two emerging forms of targeted content are expected to gain traction: Virtual Product Placements (VPPs) and Virtual Spokespersons (VSPs), detailed below.

**Virtual Product Placements** (VPPs) are promotional objects or experiences that are incorporated naturally into a user's ambient reality (virtual or augmented) on behalf of a paying sponsor. VPPs can be presented simultaneously to large groups of users, but with the advent of generative AI, these experiences are likely to be personalized at an individual level. In other words, VPPs will be custom-crafted for specific people at specific times and places to achieve specific influence goals. For example, if a user is profiled as a sports fan of a particular age, gender, and income, they might have a real-world experience in which they see someone walking past them on the street wearing a jersey for their favorite team that advertises a sports bar a few blocks ahead. Because this is a *targeted VPP*, other people around them might not see the same content, instead being targeted with very different customized promotional experiences [29].

With AR technology rapidly advancing, VPPs could soon be integrated into our surroundings so naturally that users could easily mistake them for authentic objects, activities, or even people in their proximity. If consumers can't distinguish between true experiences and targeted content, marketing could become highly predatory. This means that as digital marketing shifts from flat ads and videos to *promotionally altered environments and experiences,* consumers need new protections [30]. A simple but powerful protection would be to require that all Virtual Product Placements look *visually distinct* from their corresponding real-world artifacts. For example, if a digital product is placed into natural surroundings as promotional content, that virtual product could glow, shimmer, or be highlighted in some other visually noticeable way so it cannot be confused with authentic elements.

**Virtual Spokespersons (VSPs)** are simulated humans or other characters that are introduced into immersive environments to verbally convey targeted content by engaging a user in interactive promotional conversation [29, 30]. As AR technology becomes commonplace, it is likely that physical establishments such as retail stores and restaurants will be serviced by AI-powered virtual salespeople, virtual servers, and other virtual representatives that supplement the human staff. Similarly, when users go to the gym, they will likely be supported by a VSP virtual trainer giving them guidance and encouragement. While this seemed like pure science fiction just a few years ago, LLMs and generative avatars have made VSPs not just viable, but likely to be deployed widely in immersive platforms, motivated by the cost-savings associated with replacing human employees with interactive AI agents [14].

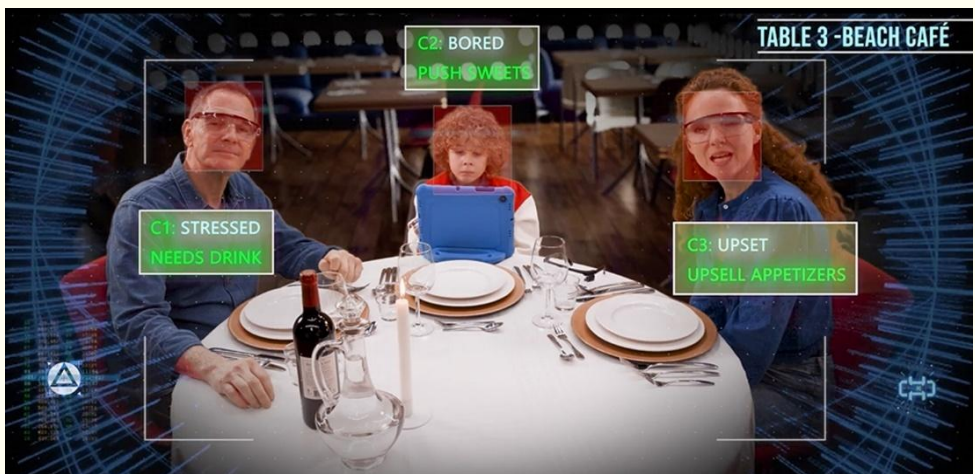

Fig 4. Image from short film "Privacy Lost" about AI manipulation dangers [31]

As shown in Figure 4 above, the use of AI and AR in retail establishments was depicted in the award-winning 2023 short film "[Privacy Lost](#)" that aimed to educate policymakers and regulators on the unique manipulative dangers of AI-powered augmented reality glasses. In the film, a family sits in a restaurant wearing fictional AI glasses called "carbons" that enable them to be serviced by VSP servers. What is potentially most interesting about the film is that member of the family – husband, wife, and son – each see a different-looking server, optimized for their perspective. And the server, by reading their facial expressions and other emotional cues, is able to upsell each family member with AI-powered manipulative skill ([video link](#)).

Another instructive short film is "Hyper-Reality" (2014) that depicts the daily life of a headset wearer in the first person as she navigates a busy city. What is most memorable about the depiction is how visually overwhelming promotional content can become if advertising is unrestrained via smart glasses or AR ([video link](#)). With the addition of AI agents that target users through interactive dialog, it is likely that promotional content deployed through AR systems will become less of a visual competition for our eyes, and more of a cognitive competition for our attention. As every salesperson knows, the best way to persuade a customer is not to hand them a brochure or have them watch a video, it is to engage them in interactive dialog that earns their trust, discovers their reservations, and works to sway their intent [29].

While using AI agents to optimally upsell customers is not a societal threat, these same AI-powered tactics could be used to deploy optimized and customized propaganda, misinformation, and disinformation at scale [33]. For these reasons, new regulation should be considered to protect consumers from predatory AI tactics that employ targeted conversational advertising [30, 32]. At a minimum, regulators should consider requiring Virtual Spokespeople in virtual and augmented worlds to have a *distinct appearance* from authentic human representatives and they should be required to explicitly announce when they are expressing targeted promotional content and indicate who the sponsor is for that content [29, 33].

Such protections are important because these AI-driven conversational agents could easily have access to detailed profile data about the targeted user they have been assigned to engage through interactive dialog. This data is likely to include their preferences, interests, political leanings, hobbies and a historical record of their reactions in prior promotional conversations. This could enable such agents to charm users in friendly small-talk, customized to their interests and leanings, thus drawing them into a false sense of trust and familiarity. In addition, without strict regulation these AI agents are likely to also have access to realtime emotional data captured from the target user's facial expressions, vocal inflections, eye-motions, pupil dilation, and even vital signs [14, 29, 30, 43, 52]. This will enable the AI agent to adjust its conversational tactics in real-time to optimize persuasive impact.

**Intellimorphic Avatars** are adaptive AI agents that intelligently adjust their appearance, behavior, and tactics in response to individual human users in order to maximize that user's receptiveness to persuasive messaging. If not regulated, such avatars could adapt their gender, hair color, eye color, clothing style, voice, accent, speech patterns, and mannerisms to maximize influence on a targeted user based on that user's previous interactions and behaviors in combination with their real-time reactions. Even more insidiously, advertisers could deploy Intellimorphic Avatars that capture a user's facial features in real-time and incorporate hints of those features into their own facial rendering. Research has shown that incorporating a small fraction of a person's own features into an AI-generated face can significantly increase a user's feelings of trust [14]. This tendency is likely due to our implicit trust of family members who share genetic resemblance. This "familial mirroring" tactic, as well as incorporating vocal features and gestural mannerisms copied from the target user (or from a family member), could be greatly abused when VSPs are deployed for marketing, propaganda, or disinformation purposes and should be strictly outlawed [30].

Additionally, research on human behavior shows that subconscious mirroring of the verbal and non-verbal behaviors of a conversational counterpart – such as emulating their speech rhythm, gestures, intonation, and breathing patterns – can build rapport and foster feelings of connection [42]. Although using such techniques could achieve similar benefits for AI-powered agents serving as personal assistants, tutors, or other supportive roles, the potential for such "behavioral mirroring" to be abused or even weaponized is significant and should be highly regulated.

**Intellimorphic Environments** – Similar to how avatars can be deployed that intelligently adapt their appearance, voice, behaviors and mannerisms to optimize impact on target users, environments can also be intelligently adaptive to individuals in real-time. For example, when a user wearing AI-powered eyewear enters a retail store, car dealership, or other commercial setting, the environment itself could be subtly adjusted to subconsciously guide a user in a certain direction or increase the probability that the targeted user notices a particular product or service. For example, a system that is aware of a user's tastes, interests, background, or economic status could coax that user down particular aisles, maybe with more expensive products, by subtly increasing the perceived lighting in that area or by projecting simulated spotlights or other visual accents on specific physical products. Similarly, virtual blinders or barriers could be added to an environment to ensure a user does not go in that direction or does not notice things the system wants to steer them away from. In an extreme example, users walking down streets could see simulated construction fences or other barriers instead of homeless encampments or other real-world events or artifacts that an organization or municipality might want to conceal [34, 46]. Such "reality-masking" techniques, which harken back to tales of fake "Potemkin Villages" in 18th century Russia, could easily be abused as individually targeted distortions, and should be highly regulated.

## 4. The Unique Danger of AI-powered Conversational Agents

When a user is engaged in real-time dialog with an AI-powered conversational agent such as a Virtual Spokesperson (VSP) that represents a third-party, or a dedicated personal assistant such as an Electronic Life Facilitator (ELF) that follows the user through their daily life, the power balance has the potential to be highly skewed, with the human at a significant disadvantage [33]. The potential sources of power asymmetry between humans and AI agents are summarized as follows:

**Familiarity Asymmetry**: As described above, it is likely that AI agents will have access to personal data about target users, ranging from age, interests, education, and political views to their favorite teams, movies, books, and musical artists. This will empower the AI to optimize the receptibility and persuasiveness of dialog for individual targets. In contrast, the human will know nothing about the artificial entity they are conversing with. And if the AI agent is given a visual or vocal persona that represents a particular age, gender, style, or background – it is merely a façade and yet the user will subconsciously make assumptions about the AI entity and its intentions, values, or trustworthiness based on its appearance. This represents a deeply asymmetric relationship that has no equivalence when engaging with human salespeople or other representatives.

**Emotional Asymmetry**: As described above, it is likely that VSPs and other AI agents will be capable of evaluating a user's emotional state during conversations from the user's expressed dialog, vocal inflections, and facial expressions. Already Open AI's latest o4 model with multimodal video processing can assess a user's emotional state from facial expression. The human, on the other hand, will be unable to "read" their digital counterpart. That's because any emotions or vocal inflections conveyed in language, voice, or face of the simulated avatar is entirely fabricated, potentially chosen to maximize impact, and does not reflect an emotional state of the AI representative. This is a highly asymmetric relationship [33].

**Continuity Asymmetry:** It is likely that the platforms deploying conversational agents will keep track of user's reactions to targeted influence and will learn what types of persuasive tactics are most effective on individual users. The human in the loop, on the other hand, will learn nothing about the digital representatives they engage throughout their day, for they could be *digital shapeshifter*s that can assume any style, tactics, or persona. In other word, the platform generating the VSP or other conversational avatar can "get familiar" with the target human over time, learning what tactics work best, while the human may believe they are engaging a variety of different entities. This too is a deeply asymmetric scenario that has no equivalent with human-human interactions.

**Information Asymmetry:** Unlike human influencers who can make arguments and counterarguments based on human-level knowledge and experience, AI agents will be able to instantly craft a compelling conversational response that draws on a nearly infinite information pool and could easily cherry-pick points that the human could not possibly validate in real-time. In fact, an AI agent in a virtual or augmented environment could create the illusion of expertise by conversationally referencing an overwhelming barrage of factual information as a deliberate form of persuasion. This is a deeply asymmetric situation that has no direct equivalent in human-human interactions.

**Strategic Asymmetry**: Unless restricted by new policy protections, it is very likely that VSPs and other conversational agents will be trained in sales tactics, negotiation tactics, human psychology, cognitive biases, game theory, and other persuasive strategies that make them extremely effective instruments of influence, manipulation, or even deception. For example, in 2022 DeepMind developed a strategic AI system called DeepNash that learned to bluff human players in the game of Stratego and sacrifice gaming pieces for the sake of long-term victory [37]. While humans are often trained in sales tactics, it is likely that AI agents will harness a far deeper, broader, and more nuanced range of persuasive strategies than any human could master, creating a highly asymmetric power-imbalance.

**Intelligence Asymmetry**: The range of cognitive tasks in which AI agents can match or exceed human performance is expanding at an astonishing pace. Between 2012 and 2020, AI systems have surpassed human abilities on critical benchmarks for image recognition, speech recognition, reading comprehension and resolving language ambiguities. Since 2021, AI systems have exceeded average human abilities on benchmarks for text summarization, software coding, medical diagnosis, and mathematical problem solving, and can produce written documents that are perceived as indistinguishable from human generated content.

Looking forward, many researchers expect AI systems to surpass humans in benchmarks for logical reasoning and multi-step planning within the next few years. As we consider the impact of AI agents that are designed to assist us throughout our daily lives via augmented reality eyewear, we must assume they will soon be able to outreason us, out-plan us, and out-strategize us as they pursue conversational objectives. This asymmetry greatly increases the manipulative risk if AI agents are allowed to impart targeted influence without regulatory oversight.

**Authority Asymmetry**: We must consider the psychological impact on human users when it is clearly established that the AI agents whispering in our ears and guiding us through our days are significantly smarter than us on nearly every front and are likely to find better solutions to our daily problems than we could come up with ourselves and do so almost instantly. I refer to this looming intellectual power imbalance as AI systems achieving *"Cognitive Supremacy"* and I worry it will cause many individuals to blindly defer to their AI assistants rather than using their own critical thinking. Whether caused by "authority bias" or "expert bias," it is well-established that humans often defer to perceived experts, greatly increasing the potential for influence and manipulation [53, 54]. This will likely result in excessive deference to AI agents that are perceived to possess super-human intellect.

Clearly, conversational agents pose unique interactive risks that have not been faced in traditional forms of targeted media. To help the public, regulators, and policymakers appreciate the new dangers posed by interactive and adaptive forms of "generative influence," it is useful to characterize the risks using a formalization referred to as the AI Manipulation Problem, elaborated on in the next section.

## 5. Generative Influence and the AI Manipulation Problem

As described above, AI-powered mobile devices have the potential to accurately track where users go, who they are with, what they are doing, what they are looking at, what they hear around them, and even their real-time emotional reactions based on facial expressions, body posture, and physiological reactions. At the same time, AI-powered immersive display has the potential to use this information to modify what that user sees, hears, and experiences in their surroundings. When combined, these two capabilities have the potential to be an extremely powerful form of digital manipulation that could become a significant risk to human agency [32, 35, 44]. To help policymakers better appreciate how this new threat compares to prior media technologies, it's helpful to formalize the risk using some basic concepts from the engineering discipline of Control Theory (CT).

Across many engineering fields, control theory is a widely used paradigm for automatically optimizing the behavior of a target system. The target system could be the inside of a home, and a thermostat is a simple mechanism that utilizes control theory to optimize the behavior (i.e. temperature) of that home. A homeowner sets a target temperature and if the house falls below the target, a heater turns on. When the heater is on, if the house rises above the target, a heater turns off. When the system is operating as expected, the thermostat keeps the house within a narrow temperature range around the target value. This is the basic concept of a *control system*, as shown in Figure 5 below.

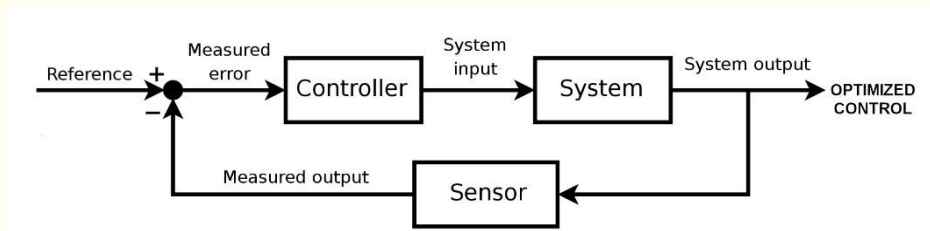

Fig 5. Classic Control System Diagram

When the schematic above is applied to a home heating unit, the block labeled *"System"* represents the house whose temperature is being controlled. The box labeled *"Sensor"* is the thermometer that detects the indoor temperature, and the box labeled *"Controller"* is the thermostat that modulates the heating unit. This creates a *feedback loop* that continually detects behaviors (e.g., temperature) and imparts influence (e.g., adjusts the heater), to guide the system towards the goal.

To appreciate the manipulation risk of AI-powered eyewear, we need to consider the case where the *System* being controlled is a human user. As shown in Figure 6 below, the input could be an **Influence Objective** from a paid sponsor, for example a request to encourage the target user to enter a particular store and buy a particular product. From an engineering perspective, this is very similar to setting the temperature on a thermostat, but in this case, it could be a complex promotional or manipulative goal with respect to the target user.

To optimally achieve its Influence Objective, the controller must access realtime sensors, which are the various onboard components that track the location and behavior of the targeted user within his or her environment. These sensors can include cameras, microphones, GPS sensors, LiDAR sensors, and other modules that monitor the user's location, orientation, actions, and dialog. These sensors could also include heartrate monitors, eye-trackers, and other more invasive elements such as skin galvanometers and EEGs. The sensor data can be processed by multimodal large language models to instantly generate a real-time assessment of that user's behaviors, intentions, and emotions with respect to their surroundings.

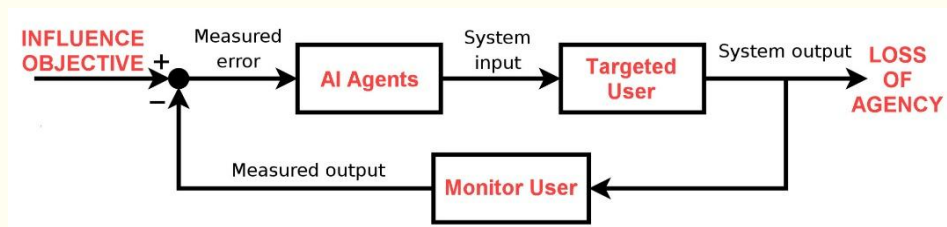

Fig 6. Control Theory representation of the AI Manipulation Problem

The system also needs a Controller that, like a thermostat, continually adjusts its tactics to guide the system towards the desired goal. In this case, the controller is a powerful Generative AI system that deploys AI agents to modulate the sights and sounds experienced by a user as they navigate their world, including targeted conversational dialog from AI-controlled avatars. And like a thermostat modulating the temperature of a home, this can be deployed as a feedback loop that continually monitors the state of the user and adjusts tactics to achieve the influence objective.

For example, imagine a user in a coffeehouse wearing AI-powered AR glasses. The target user may not notice that a simulated couple has been added to the local environment, for the controller (which has access to gaze direction of users) could easily make the couple appear at the nearby table at an instant when the user is glancing in a different direction. The couple could be rendered photorealistically and therefore perceived as authentic customers by the target user. Once the couple is created, the controller could begin imparting targeted content in a subtle way: by having the simulated couple strike up an audible conversation between themselves, their dialog being spoken within earshot of the target user. The conversation could begin as casual dialog and gradually shift to an influence objective. The controller would then monitor the user's reactions during the casual dialog to determine if and when the target user begins paying attention to the nearby couple.

Assessing attention could be as simple as detecting changes in the user's gaze or biometrics in correlation with realtime comments made by the simulated couple. For example, the user might form a slight smile when the couple says something funny, or the user's pupils might contract slightly when the couple says something provocative. Regardless of the cues, the controller could easily detect engagement, determining that the target user has begun paying attention to the overheard dialog. The controller could then drive the virtual couple to shift their conversation towards the influence objectives.

For example, the target user may be on the market for a new car. A sponsor might have purchased the right to target this user from the AR platform provider via an advertising-based business model. Without regulation, the platform provider would have no obligation to inform the user that the virtual couple are VSPs that were added to that environment for promotional purposes. Therefore, when the virtual couple begins discussing how happy they are with a recent purchase of a particular vehicle, the targeted user may believe he is overhearing the authentic dialog among customers in a coffeehouse and not a targeted influence campaign.

As the overheard conversation proceeds, the controller could monitor the target user, assessing their facial expressions, body language, pupil dilation, posture, eye motion, respiration rate and blood pressure to detect emotional reactions and adjust conversational tactics to optimize persuasive impact. For example, if the user shows increased focus when the couple begins talking about the car's engine, the controller could adapt its tactics, shifting the conversation towards vehicle performance. On the other hand, if the user's attention fades when the couple talks about the sound system of the car, the controller could adjust, shifting off that line of influence. Meanwhile, the user could be unaware that the conversation is not authentic and have no idea that it is reacting to his facial or biometric reactions.

In other words, the target user could easily become an unwilling participant in the overheard conversation, interacting through subconscious physical reactions. This may sound dystopian, but this type of feedback loop is feasible in virtual and augmented environments powered by generative AI systems. Taking this one step further, an AI-generated spokesperson could directly engage the user in interactive conversation, and using the adaptive power of AI, gradually talk them into accepting a targeted influence objective, whether it be buying a specific product or service, embracing a marketing message or propaganda, falling for a scam, or believing misinformation or disinformation. In all such cases, it's the interactive and adaptive nature of generative influence that creates the unique risk. This is especially true when a real-time control system is established around the user, creating a feedback loop that can optimize persuasive tactics in real-time.

Thus, **The AI Manipulation Problem** refers to the unique manipulative risk to human agency when an AI-powered system establishes a feedback loop around a target user and performs the following steps:

(i)     Impart targeted influence on an individual user,
(ii)    Detect that user's reaction to the imparted influence,
(iii)   Adjust influence tactics in response to the user's detected reaction,
(iv)    Repeat steps i, ii, and iii to optimally achieve influence objective.

These are simple steps, much like a thermostat controlling the temperature in a house, and yet it could result in the most effective form of targeted manipulation ever developed. The danger is significantly amplified when the generative AI systems are placed onboard devices such as AI eyewear that can track a target user's location, behavior, and emotions, while maintaining a context-aware model of that user's environment. Unless regulated, this unique convergence of context-aware generative AI, mobile computing, and augmented reality could compromise human agency at scale by enabling individually customized targeted manipulation and persuasion with unprecedented levels of effectiveness [35, 51].

## 5. Regulatory Protections

As it becomes increasingly likely that AI-powered augmented reality will be central to our digital lives, we must consider the need for regulation. With respect to the AI manipulation problem described above, the most direct regulatory approach would be to ban the use of *real-time feedback loops* around human users in which AI agents work to optimize their persuasive impact. [33, 35]. In addition, we must consider the risks of behavioral tracking, emotional profiling, and virtual product placements:

**Behavioral Tracking**: To enable real-time AR experiences through AI powered eyewear, platform providers must be able to monitor user location and behaviors, but they do not need to retain this data and should not be allowed to. Limiting data storage would prevent invasive behavioral profiling of user activities [7, 29].

**Emotional Profiling**: Already, many immersive devices enable real-time tracking of facial expressions, eye motion, posture, and vital signs to determine a user's real-time emotional and physiological states. Without regulation, emotional profiling could be performed to optimize manipulation by AI agents. Limits should be placed on the tracking and usage of emotional data, and users should be clearly informed whenever their emotional or physiological reactions are monitored [7].

**Virtual Product Placements**: As augmented reality becomes commonplace, advertising will include promotional products and services seamlessly integrated into our daily lives, potentially becoming indistinguishable from real experiences. This can easily be used to impart subconscious influence, affecting consumer sentiments and even political views. Regulations should mandate clear visual indicators for virtual product placements in a user's environment [29].

## 6. Conclusions

Technologies for augmenting our perception of reality have been in development for decades, starting from the early head-up displays of the 1950s and 1960s, to the far more immersive and interactive products of today [3, 9, 50]. During most of that period, little concern has been given to the subtle manipulative risks of targeted AR content. In recent years, however, the rapid advancement of augmented reality, generative AI, and conversational AI agents have created significant new threat vectors for immersive influence and targeted manipulation [29-33]. This is especially significant for "true AR systems" that foster distal attribution and thus reduce the mental distinctions between what is real and what is digital [3-6, 14].

As described above, we must consider regulatory protections, such as banning the use of *feedback loops* that enable AI agents to optimize their influence during real-time interactions [33, 35]. In addition, regulators should ban the appropriation of a user's own facial features, vocal qualities, gestural mannerisms and speech patterns in the depiction of conversational agents. Also, regulators should require all promotional alterations to a user's environment be clearly distinguishable from authentic elements [7, 14, 29]. Finally, regulators must be aware that the manipulative risks posed by AI agents could increase dramatically as AI systems achieve real or perceived *cognitive supremacy* with respect to human users.

ABOUT THE AUTHOR

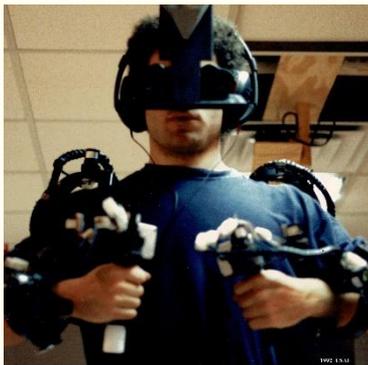

Louis Rosenberg developing AR in 1992.

Louis Rosenberg, PhD is an early pioneer of augmented reality and a longtime AI researcher. He developed the first interactive augmented reality system in 1992 at Air Force Research Laboratory (AFRL) and published the first papers showing that AR could enhance human performance in real-world tasks. He founded the early VR company Immersion Corporation in 1993, the early AR company Outland Research in 2004, and the artificial intelligence company Unanimous AI in 2014. He earned his PhD from Stanford University, was a tenured professor at California State University (Cal Poly), and has been awarded over 300 patents for his work in VR, AR, AI, and haptics.